Visualization of Organ Movements Using Automatic Region Segmentation of Swallowing CT

Yukihiro MICHIWAKI[1,2], Takahiro KIKUCHI[3], Takashi Ijiri[4], Yoko Inamoto[5], Hiroshi MORIYA[6], Takumi OGAWA[7], Ryota Nakatani[8], Yuto Masaki[8], Yoshito Otake[8], Yoshinobu SATO[8]

1) Toho University, School of Medicine, 2) Michiwaki Kenkyusho, Co., Ltd.
3) R&D Division, Meiji Co., Ltd., 4) Shibaura Institute of Technology, 5) Faculty of Rehabilitation, School of Health Sciences, Fujita Health University, 6) Diagnostic Imaging Center, Ohara General Hospital, 7) Department of Fixed Prosthodontics, Tsurumi University School of Dental Medicine, 8) Nara Institute of Science and Technology, Division of Information Science

**Introduction**

Swallowing is one of the "invisible phenomena." Visualizing "invisible phenomena" is extremely useful for deepening and standardizing understanding. The quality of visualization can be broadly categorized into two types: visualization that can be seen if one has prior assumptions and visualization that is undeniable to everyone. The former varies greatly depending on the quality of the prior assumptions, meaning that there are significant differences between observers. Prior assumptions can also be referred to as conjecture. The latter leaves little room for conjecture and has minimal differences between observers. When the goal is to deepen and standardize understanding, the quality of visualization should ideally approach the latter.

Methods of visualizing swallowing movements include medical imaging such as videoendoscopy (VE), ultrasonography (US), videofluorography (VF), and four-dimensional computed tomography (4D-CT) for swallowing (hereafter referred to as swallowing CT). VE provides clear pharyngeal images, however, the organs are not visible during swallowing due to whiteout[1]. Ultrasound images infer tongue and pharyngeal wall movement from a single cross-sectional image[2]. VF, which involves swallowing a contrast agent, excels in evaluating aspiration and residue[3]. However, explaining the three-dimensional shape and movement asymmetry of organs using VF requires considerable conjecture. Swallowing CT provides three-dimensional and temporal information on boluses and organs[4]. However, due to the limitations of spatial and temporal resolution of imaging devices, the accuracy of information decreases at high-speed movement frames, sometimes resulting in artifacts such as double images of the hyoid bone

(Fig.1). At present, there is no medical technology that can visualize organ movement at a level that is undeniable to everyone.

Meanwhile, advances in deep learning have improved the accuracy of artificial intelligence (AI) in automatically recognizing organs in medical images[5].

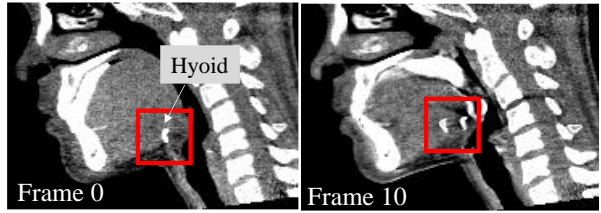

Fig. 1. Examples of motion blur. Before swallowing begins (Frame 0) and during swallowing (Frame 10). During swallowing, two hyoid bones are visible.

The basic concept of automatic recognition is to explicitly delineate boundaries for regions of interest, known as segmentation. In swallowing-related research, Zhang, Z. used deep learning-based object detection methods such as SSD (single shot multibox detector) to track the hyoid bone in VF images[6]. Caliskan[7] and Ariji[8] performed automatic segmentation of boluses and evaluated concordance using Dice coefficients. However, no reports exist on AI development for segmenting boluses, hyoid bones, tongue, and soft palate using swallowing CT.

The purpose of this study was to develop AI using the 3D convolutional model of nnU-Net[9] to perform automatic segmentation of organ three-dimensional shapes. This report also demonstrated that this AI provided relatively clear visualization of organ movements during swallowing.

**Materials and Methods**

**1. Materials**

The research materials consist of 4D-CT images taken during swallowing or chewing and swallowing. The imaging equipment used was a 320-row area detector CT (Aquilion ONE global standard edition; Canon Medical Systems), with a rotation time of 0.35 seconds, continuous imaging for 2-3 seconds, half reconstruction, an image acquisition time of approximately 0.2 seconds per frame, and a frame rate of 10 fps.

The study subjects were five healthy adults (ages 24-61, three males and two females). The bolus used was a barium sulfate solution with a thickening agent (5% w/v), with volumes ranging from 4 to 10 ml. The image size was 512x512x320, and the number of frames and voxel sizes varied depending on the case (Tab.1). The swallowing style was a command swallow.

| Data of 4D-CT | |
|---|---|
| Nr. of subjects | 5 (healthy volunteers) |
| Ages and sex | 24～61 years old, 3 males and 2 females |
| Nr of frames | 22~31 (129 in total) |
| Size of images | 512x512x320 |
| Voxel size (mm) | (0.468~0.625)x(0.468~0.625)x0.500 |

Tab. 1. Swallowing CT dataset

**2. Methods**

**1) Development of segmentation AI using supervised learning**

(1) Creation of ground truth data

Ground truth data was created from five swallowing CT cases. The total number of frames was

129, corresponding to 129 volumetric images in conventional CT. The segmented regions included the tongue and soft palate for soft tissues, bones such as facial bones, mandible, cervical vertebrae and hyoid, cartilages as thyroid cartilage and epiglottis, and the bolus.

Ground truth data was created through manual segmentation using the RoiPainter4D software[10] which was developed for ourself. Different segmentation methods were used for each region. Bone and cartilage were segmented using rigid body tracking method, while the bolus was segmented using a range-restricted region expansion method. The tongue and soft palate were segmented using the free-form deformation (FFD) method. This approach involves creating a standard model and deformation cage for each tissue and editing cage vertices to deform the standard model to match the CT image.

The labeling of organs other than the tongue was conducted based on anatomical knowledge. The deep boundary of the intrinsic tongue muscle is unclear, so in this study, it was defined as a curved surface connecting the floor of the mouth mucosa and the lesser horn of the hyoid bone.

The annotators included three dentists and one engineering graduate. Final verification of the annotations was conducted collaboratively by one dentist and one engineer.

(2) Development of segmentation AI

The AI model used was the 3D convolutional model of nnU-Net, which excels in labeling each pixel and automatically generating segmentation architectures tailored to the dataset.

The AI training and evaluation method employed was leave-one-out cross-validation. Four out of the five cases were used for training, and the remaining case was used as test data, repeated for each case. The number of epochs was set to 100.

## 2) AI Accuracy Verification

(1) Calculation of Dice coefficients

Dice coefficients[11] were used to evaluate the segmentation accuracy of the AI. The Dice coefficient quantifies the overlap ratio between the ground truth data and the AI-segmented regions, with a value of 1 indicating perfect agreement and 0 indicating no overlap. The Dice coefficient was calculated for each segmented region across 129 frames.

(2) Analysis of regions with insufficient Dice coefficient values

Regions with low median Dice coefficients or high standard deviations were analyzed individually by comparing swallowing CT images, ground truth data, and AI segmentation data to investigate potential reasons for inadequate Dice coefficient values.

## 3) Automatic Region Segmentation and Qualitative Evaluation of Chewing and Swallowing 4D-CT

To evaluate the practical applicability of AI, region segmentation and qualitative evaluation were conducted using a dataset separate from the five cases used for ground truth data creation. The AI model was trained using all five cases with nnU-Net. The study materials included chewing and swallowing 4D-CT scans. The subject of the chewing and swallowing CT was one case, using 5g of contrast agent-containing gummy food, with 57 frames. The image size was the same as the swallowing CT, with a voxel size of 0.468x0.468x0.500 mm³. The swallowing method was free swallowing.

Although swallowing and chewing-swallowing are different movements, they involve common regions, making AI accuracy verification appropriate. While the training data consisted of a maximum of 31 frames, the chewing-swallowing CT had 57 frames. For ease of comparison, this report presents the region segmentation results for the latter 30 frames corresponding to the swallowing phase.

**Results**

**1. Distribution of Dice Coefficients for Segmented Organs**

The median Dice coefficient for the food bolus was 0.80, with a standard deviation of 0.31. The Dice coefficient values for the facial bones, mandible, and cervical vertebrae had medians of 0.8 or higher, with standard deviations below 0.1. The median Dice coefficient for the hyoid bone was 0.78, with a standard deviation of 0.14. The median values for the thyroid cartilage and epiglottis were 0.59 and 0.32, respectively, with standard deviations of 0.12 and 0.24. The epiglottis had the lowest Dice coefficient among the segmented regions.

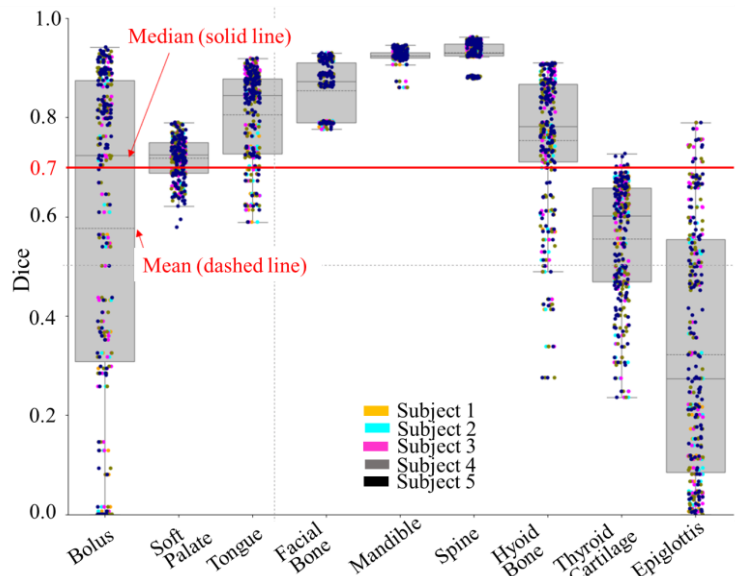

Fig. 2. Box plot of Dice coefficients for food boluses and organs. A Dice coefficient of 0.7 is considered a practical guideline.

For soft tissues, the tongue had a median Dice coefficient of 0.85 and a standard deviation of 0.08. The soft palate had a median Dice coefficient of 0.72, with a standard deviation of 0.05. The median and standard deviation of Dice coefficients varied by tissue type (Fig. 2).

**2) Individual Analysis of Areas with Insufficient Dice Coefficient Values**

Regions with low median Dice coefficients or large standard deviations were analyzed individually. For example, although the food bolus had a favorable median value, its standard

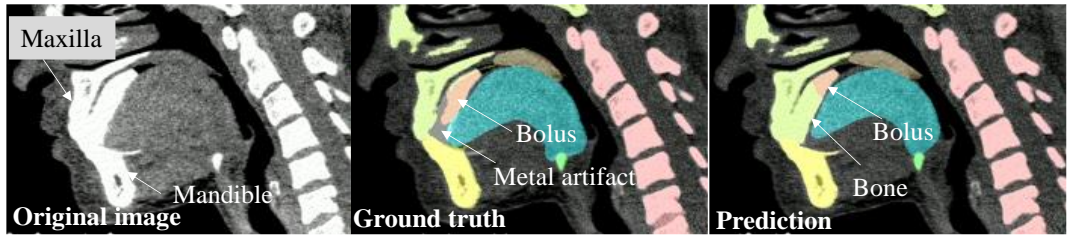

Fig. 3 Comparison of swallowing CT (left), ground truth data (center), and AI prediction (right) for the mandible, tongue, and soft palate. The three are in good agreement regarding the mandible, tongue, and soft palate. The lower accuracy of AI prediction for the bolus is considered to be due to metal artifacts.

deviation was large due to metal artifacts caused by dental crowns (Fig. 3).

For rapidly moving time points, the hyoid bone showed deformation and defects in the greater horn. The thyroid cartilage exhibited wedge-shaped defects, holes, and deformations at the upper edge (Fig. 4). The epiglottis showed tip defects during inversion.

### 3) Automatic Region Segmentation of Chewing and Swallowing 4D-CT

AI-based region segmentation of the chewing-swallowing 4D-CT did not show significant misrecognition for the facial and mandibular bones or the tongue. Although the smoothness of the mucosal surface depiction for the tongue and soft palate was insufficient, no errors were observed in their morphology. For the hyoid bone, thyroid cartilage, and epiglottis, parts of the greater horn of the hyoid bone, the superior horn of the thyroid cartilage, and the tip of the epiglottis were not depicted during rapid movements. Thus, the AI's segmentation performance for new cases was consistent with the Dice coefficient values calculated from the training data (Fig. 5).

### Discussion
### 1. Significance and Challenges of AI-Based

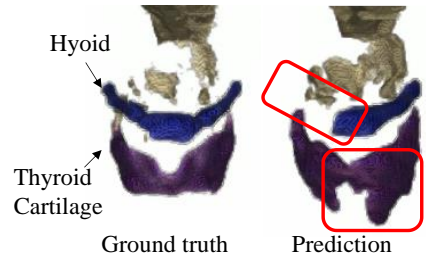

Fig. 4. Ground truth data (left) and AI prediction (right) during swallowing (Frame 9). In the AI prediction, the right greater horn of the hyoid bone is missing, there is a large defect in the midline of the thyroid cartilage, and an excess region is observed on the left side.

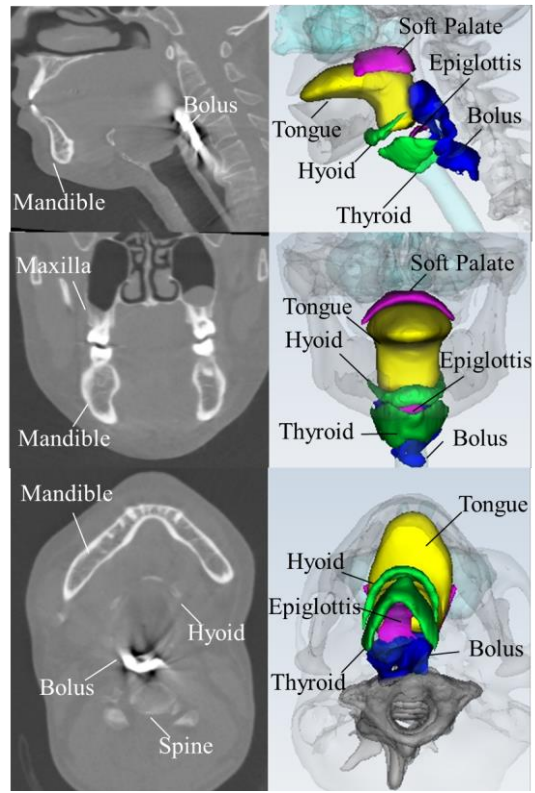

Fig. 5. Swallowing CT during the pharyngeal phase (left) and the image after AI-based segmentation (right).

**Automatic Region Segmentation**

The primary significance of AI-based automatic region segmentation is time reduction. Manual segmentation of the tongue, soft palate, thyroid cartilage, and epiglottis using RoiPainter4D took approximately two months per case. In contrast, the AI model completed the same task within about 60 minutes, fully automated without manual intervention. This AI significantly reduces the labor and time required for segmentation, enabling the processing of numerous cases in the future. The main challenge is improving AI accuracy. The Dice coefficient distribution is influenced by the clarity of the original swallowing CT images. The highest Dice coefficients were observed for the mandible, facial bones, and cervical vertebrae due to their large size, high density, and clear contrast with surrounding tissues, as well as minimal movement during swallowing.

Tissues with low clarity in swallowing CT include the hyoid bone, thyroid cartilage, and epiglottis. The hyoid bone exhibits varying degrees of ossification between the body and the lesser and greater horns. Additionally, the lesser horn, being a small projection, and the greater horn, being slender and rod-shaped, both tend to become unclear in swallowing CT during rapid movements. The thyroid cartilage, being thin and plate-like with low and uneven ossification, often appears unclear at the edges and may show partial defects even at rest. During rapid movements, it becomes even more challenging to identify the superior horn of the thyroid cartilage. The epiglottis, being a thin, plate-like structure with a lower degree of ossification than the thyroid cartilage, moves at a rapid speed during inversion in swallowing. As a result, it often appears unclear in swallowing CT, and its inversion is rarely discernible.

Despite these challenges, swallowing CT remains the only medical imaging technique that captures 3D and temporal information of these structures. Improving visualization quality through noise reduction and segmentation method enhancements is valuable.

It still remains challenging to eliminate image noise such as metal artifacts and motion blur associated with CT imaging. We consider a region segmentation method leveraging the material characteristics of bones and cartilage to be effective. Since bones and cartilage do not experience significant defects or deformation during motion, their three-dimensional structure at rest can be set as the default, with positional changes tracked to account for motion.

For the tongue and soft palate, their boundaries are clear at rest due to the presence of an air layer with significantly different CT values. However, during swallowing, the mucosal surfaces come into contact, and the rapid and extensive movements render the swallowing CT images unclear. Despite this, the Dice coefficients for both regions exceeded 0.7, with small standard deviations. This is likely because the tongue occupies a large portion of the oral cavity, and in the current cases, no abnormalities in tongue morphology or movement were present, making its structure easier to recognize. Additionally, since the Dice coefficient evaluates the overlap ratio, regions with the same mismatch area but a larger total area may yield smaller Dice coefficient values.

Upon closer examination of individual tongue images, minor deformations and unnatural irregularities on the dorsal surface of the tongue were observed. To improve AI's region segmentation accuracy, we aim to increase the amount of ground truth data, include cases demonstrating abnormalities in tongue morphology and movement, and explore evaluation methods beyond Dice coefficients.

For the food bolus, the standard deviation of the Dice coefficient was large. Individual image analyses revealed that factors such as the influence of metal artifacts, confusion with the hyoid bone or thyroid cartilage, and leaks outside the digestive tract due to motion blur rendered the food bolus boundaries unclear. To improve the Dice coefficient for the food bolus, methods such as segmenting the hyoid bone or thyroid cartilage prior to food bolus segmentation and defining the food bolus pathway based on the digestive and respiratory tracts may be useful.

While such studies are expected to improve AI's region segmentation accuracy, the risk of misrecognition will always persist. Therefore, it is essential to develop tools that efficiently correct AI-generated region segmentation results.

**2. Future Challenges**

Region segmentation clarifies the boundaries of the regions of interest, enhancing visualization quality. In this study, the segmentation of the three-dimensional shapes of organs and the food bolus enabled the visualization of organ movements and food bolus transit, which were unclear with swallowing CT alone. On the other hand, new challenges emerged from visualization.

For example, during vertical movement of the hyoid bone, some cases exhibited asymmetry in the movement of the greater horns of the hyoid bone. Anatomically, both the morphology of the hyoid bone and the arrangement of the suprahyoid muscle group are not symmetrical. Additionally, neural innervation is unilateral. From a functional anatomical perspective, it is possible for asymmetry to occur in the vertical movement of the greater horns during swallowing, but this had not been demonstrated using medical imaging before.

Additionally, there were cases where the greater horns of the hyoid bone appeared to move closer together during the swallowing motion. The lesser and greater horns of the hyoid bone are attached to the suprahyoid muscles as well as the middle pharyngeal constrictor muscle. When the middle pharyngeal constrictor contracts, it acts to narrow the pharyngeal cavity. This muscular force may pull the greater horns of the hyoid bone inward, overcoming the low elasticity of the hyoid bone.

Artifacts such as motion blur also affect the accuracy of swallowing CT. Improving the temporal resolution of CT equipment and reducing artifacts, along with enhancing AI segmentation accuracy, are essential areas for future research.

AI-driven visualization is expected to contribute not only to advancing the analysis of organ movements during swallowing but also to improving the accuracy of swallowing CT by clarifying its current precision levels.

This study was approved by the Clinical Research Ethics Committee of Musashino Red Cross Hospital (Approval No. 129).

**Research Funding and Conflicts of Interest**
This study was conducted with funding from the Ministry of Education, Culture, Sports, Science and Technology's Scientific Research Fund, Basic B (Project Number: 23K28478), and a joint research commission from Meiji Co., Ltd.

**Literature**
1. Chiaki Susa, Hitoshi Kagaya, Eiichi Saitoh, Mikoto Baba, Daisuke Kanamori, Shinya Mikushi, Kenichiro Ozaki, Hiroshi Uematsu, Shunsuke Minakuchi. Classification of sequential swallowing types using videoendoscopy with high reproducibility and reliability. Am J Phys Med Rehabil. 2015. 94(1):38-43. doi: 10.1097/PHM.0000000000000144.
2. Keisuke Maeda, Motoomi Nagasaka, Ayano Nagano, Shinsuke Nagami, Kakeru Hashimoto, Masaki Kamiya, Yuto Masuda, Kenichi Ozaki, Koki Kawamura：Ultrasonography for Eating and Swallowing Assessment: A Narrative Review of Integrated Insights for Noninvasive Clinical Practice. Nutrients. 2023. 12;15(16):3560. doi: 10.3390/nu15163560
3. Jeri A. Logemann, Alfred W. Rademaker, Barbara Roa Pauloski, Yukio Ohmae, Peter J. Kahrilas. Normal Swallowing Physiology as Viewed by Videofluoroscopy and Videoendoscopy. Folia Phoniatr Logop. 1998. 50 (6): 311–319. https://doi.org/10.1159/000021473
4. H. Bayona, Y. Inamoto, E. Saitoh, K, Aihara, M. Kobayashi and Y. Otaka. Prediction of Pharyngeal 3D Volume Using 2D Lateral Area. Measurements During Swallowing, Dysphagia, 2024, doi.org/10.1007/s00455-023-10659-x
5. Hiasa Y, Otake Y, Takao M, Ogawa T, Sugano N, Sato Y. Automated muscle segmentation from clinical CT using Bayesian U-Net for personalized musculoskeletal modeling. IEEE transactions on medical imaging. 2019. 10;39(4):1030-40.
6. Zhang, Z., Coyle, J. L. & Sejdić, E. Automatic hyoid bone detection in fluoroscopic images using deep learning. Sci. Rep. 2018. 8, 12310
7. Caliskan, H., Mahoney, A. S., Coyle, J. L. & Sejdic, E. Automated bolus detection in videofluoroscopic images of swallowing using mask-RCNN. Annu. Int. Conf. IEEE Eng. Med. Biol. Soc. 2020, 2173–2177
8. Yoshiko Ariji, Masakazu Gotoh, Motoki Fukuda, Satoshi Watanabe, Toru Nagao, Akitoshi Katsumata & Eiichiro Ariji. A preliminary deep learning study on automatic segmentation of contrast-enhanced bolus in videofluorography of swallowing. Nature Scientific Reports. 2022. 12, 18754　https://doi.org/10.1038/s41598-022-21530-8

9. Isensee, F., Jaeger, P. F., Kohl, S. A., Petersen, J., & Maier-Hein, K. H. nnU-Net: a self-configuring method for deep learning-based biomedical image segmentation. Nature methods, 2021.18(2); 203-211.

10. Kimura, Y., Ijiri, T., Inamoto, Y., Hashimoto, T. and Michiwaki, Y.: Interactive segmentation with curve-based template deformation for spatiotemporal computed tomography of swallowing motion, *PLOS ONE*, 2024. 19(10); 1–16, DOI: 10.1371/journal.pone.0309379.

11.Zou KH, Warfield SK, Bharatha A, Tempany CM, Kaus MR, Haker SJ, Wells III WM, Jolesz FA, Kikinis R. Statistical validation of image segmentation quality based on a spatial overlap index1: scientific reports. Academic radiology. 2004. 11;178-89. https://doi.org/10.1016/S1076-6332(03)00671-8